\documentclass[aip,apl,bmf,sd,rsi,amsmath,amssymb,reprint,author-year,author-numerical,floatfix]{revtex4-1}
\usepackage{graphicx}
\usepackage{bm}
\usepackage{gensymb}
\usepackage{dcolumn}
\usepackage[dvipsnames]{xcolor}
\usepackage{graphicx}
\usepackage{dcolumn}
\usepackage{bm}
\usepackage{soul}
\usepackage{float}
\usepackage[hidelinks=true]{hyperref}
\usepackage[hyphenbreaks]{breakurl}
\hypersetup{
  colorlinks   = true, 
  urlcolor     = blue, 
  linkcolor    = blue, 
  citecolor    = blue 
}
\begin{document}

\title{Magnetic field induced arrested state and observation of spontaneous anomalous Hall effect in TbMn$_6$Sn$_6$.}

\author{Tamali Roy}
\author{Prasanta Chowdhury}
\author{Mohamad Numan}
\author{Saurav Giri}
\author{Subham Majumdar}
\email{sspsm2@iacs.res.in}
\affiliation{School of Physical Sciences, Indian Association for the Cultivation of Science, 2A \& B Raja S. C. Mullick Road, Jadavpur, Kolkata 700 032, India}
\author{Sanat Kumar Adhikari}
\author{Souvik Chatterjee}
\affiliation{UGC-DAE Consortium for Scientific Research, Kolkata Centre, Sector III, LB-8, Salt Lake, Kolkata 700106, India}

\begin{abstract}
The quasi two-dimensional kagome ferrimagnet TbMn$_6$Sn$_6$ is investigated for thermo-remanent magnetization and Hall effects. On cooling under a moderate magnetic field, the sample attains a magnetization value close to the saturation magnetization. Upon heating in a very small magnetic field, the sample continues to maintain the large value of magnetization, which eventually diminishes distinctly  at around 200 K manifesting an ultrasharp jump. A similar feature is also observed in the Hall resistivity, which holds its saturation value when heated back in zero field  after being field-cooled. The ultrasharp jump in magnetization is also get reflected in our Hall data. The observed data is exotic and can be rooted to the large anisotropy and the strong exchange interaction.    

\end{abstract}

\maketitle

Quasi two-dimensional kagome spin systems, in the presence of geometrical frustration, strong spin-orbit coupling, and electron-electron correlation, can host magnetic topological phases such as spin liquid, flat electronic band, Dirac and Weyl semimetal, alongside manifesting some unique magnetic behaviors. Recently, they are emerging as important materials for spintronics and quantum information technology. TbMn$_6$Sn$_6$ is one such metallic kagome magnet belonging to the well-known RMn$_6$Sn$_6$ (R= rare earth element) family which shows number of interesting magnetic and electronic properties ~\cite{RMS_mag,AHE_RMS_PRL,THE_YMS_1,THE_YMS_2}.

\par
TbMn$_6$Sn$_6$ crystallizes in a hexagonal $P6/mmm$ structure, which consists of a clean Mn kagome layer alongside alternate Tb and Sn layers between them~\cite{Riberolles2022}. It is a collinear ferrimagnet (FIM) with a relatively high Curie temperature ($T_c$= 423 K), where the spins lie perpendicular to the $c$ axis of the crystal. Upon cooling, it undergoes a drastic spin reorientation (SR) transition at about $T_{sr} \approx$ 310 K where all the spins align themselves along the $c$-axis ~\cite{RMS_singlecrystal_Jmmm,RMS_neutron,SR_transition}. The intra-layer Tb-Tb  and Mn-Mn interactions are ferromagnetic whereas the inter-layer Tb-Mn interaction is anti-ferromagnetic~\cite{Riberolles2022}, giving rise to the FIM ordering with net saturation moment 5.8 $\mu_{B}$ around 200 K~\cite{TMS_chern_Topology}. Mn has an easy plane anisotropy and Tb has a uniaxial anisotropy, and their competition causes the SR transition ~\cite{anisotrpy_1,anisotrpy_2,SR_transition_2}. 
   
\par
TbMn$_6$Sn$_6$ shows large intrinsic anomalous Hall conductivity with a value $\sigma^A_{xy} \approx$ 130 Scm$^{-1}$, which  comes from the Berry curvature of the electronic bands~\cite{AHE_Gao}. This compound also shows anomalous Nernst and anomalous thermal Hall conductivity~\cite{ANE_ATHE_AHE_nat}, which also bear the same origin. TbMn$_6$Sn$_6$ has been found to be a topological magnet, as it hosts Chern-gapped Dirac fermions~\cite{TMS_chern_Topology}.
\par
The fascinating magnetic and electronic properties of TbMn$_6$Sn$_6$ are linked to the  kagome (Mn) and triangular (Tb) layers along with the intricate magnetic anisotropy. Despite extensive magnetic studies on the material~\cite{net_moment,anisotrpy_1,SR_transition,RMS_singlecrystal_Jmmm,RMS_neutron,RMS_mag}, there is little study on the evolution of remanent magnetization in the compound. Here, we systematically study the magnetization ($M$) and Hall resistivity ($\rho_{xy}$) on the field-cooled state of the sample. Our study indicates the presence of strong thermo-remanent $M$, which also gives rise to spontaneous $\rho_{xy}$ in the absence of any applied magnetic field ($H$). The thermo-remanence eventually collapses with an ultrasharp jump when the sample is heated above 200 K. 
\par
The present study was performed on high-quality single crystals (residual resistivity ratio close to 100) synthesized by the Sn flux method as described in Appendix A. We prepared two different batches of samples, namely, B1 and B2 with the almost same heat treatment. The crystals from B1 and B2 have identical magnetic properties, and in the main text, we provide the data mostly from B1. 

\begin{figure}[hbt]
	\centering
	\includegraphics[width = 7 cm]{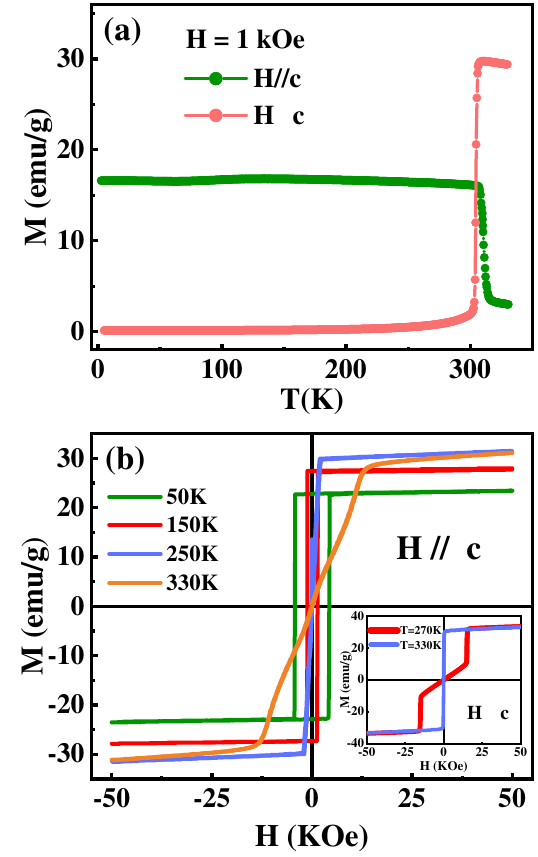}
	\caption{(a) Temperature dependence of magnetization along (green) and perpendicular (red) to the $c$-axis respectively. (b) Field dependence of magnetization at different temperatures along the $c$-axis. The inset shows the M-H curve measured at 270 and 330 K with $H\perp c$.}
	\label{fig:MAG}
\end{figure}

\begin{figure}[hbt]
	\centering
	\includegraphics[width = 8 cm]{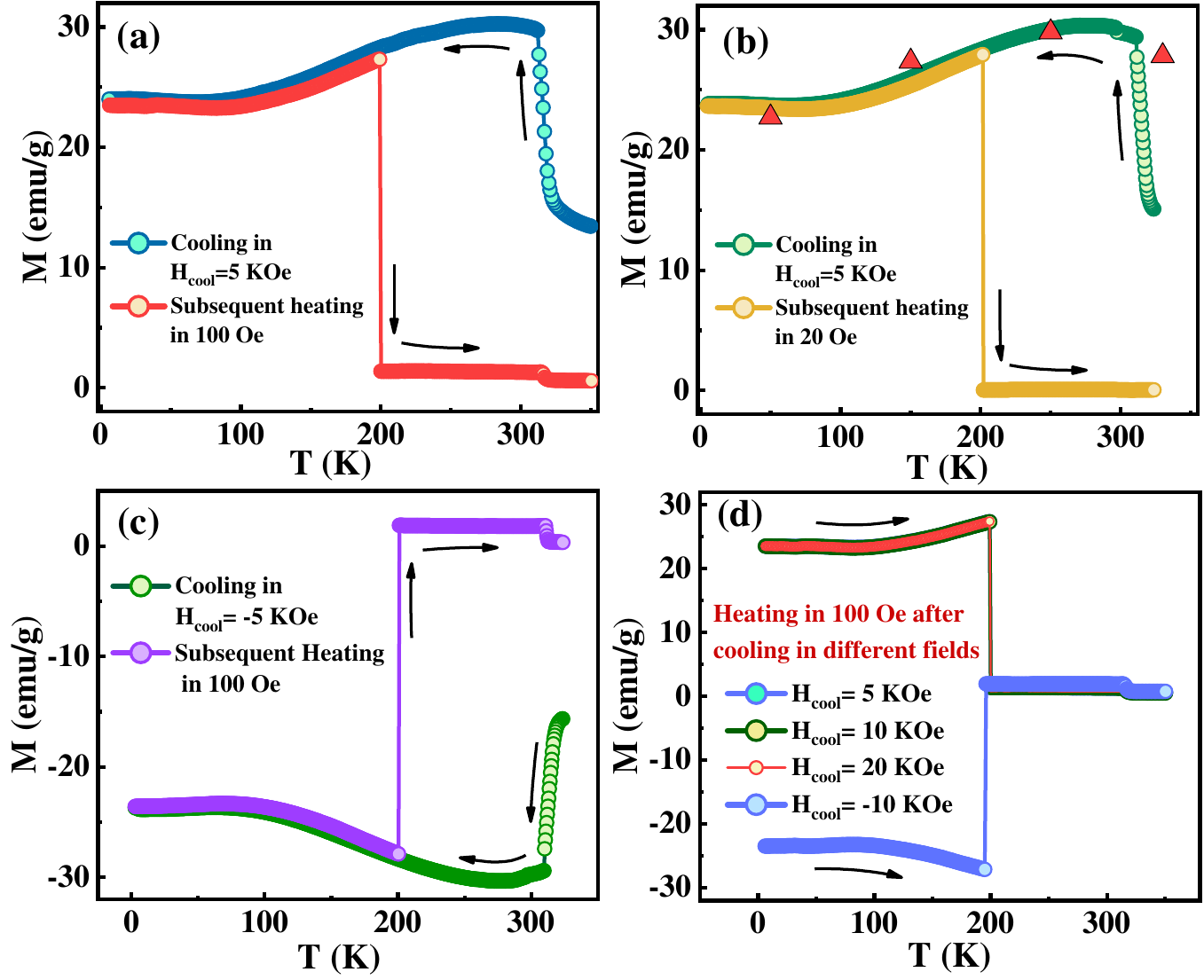}
	\caption{Temperature dependence of magnetization while cooling and heating in different fields with $ H \parallel c$ axis (a) Cooling in 5 kOe and heating in 100 Oe  (b) Cooling in 5 kOe field and heating in 20 Oe. The scattered red triangles are the saturation values of $M$  at various $T$ obtained from the isothermal $M$-$H$ curves [Fig.~\ref{fig:MAG} (b)]. (c)  Cooling in -5 kOe field and heating in 100 Oe  (d) Heating in 100 Oe after cooling in different applied fields i.e. 5, 10, 20, and -10 kOe.}
	\label{fig:TRM_mag}
\end{figure}

\begin{figure}[hbt]
	\centering
	\includegraphics[width = 8 cm]{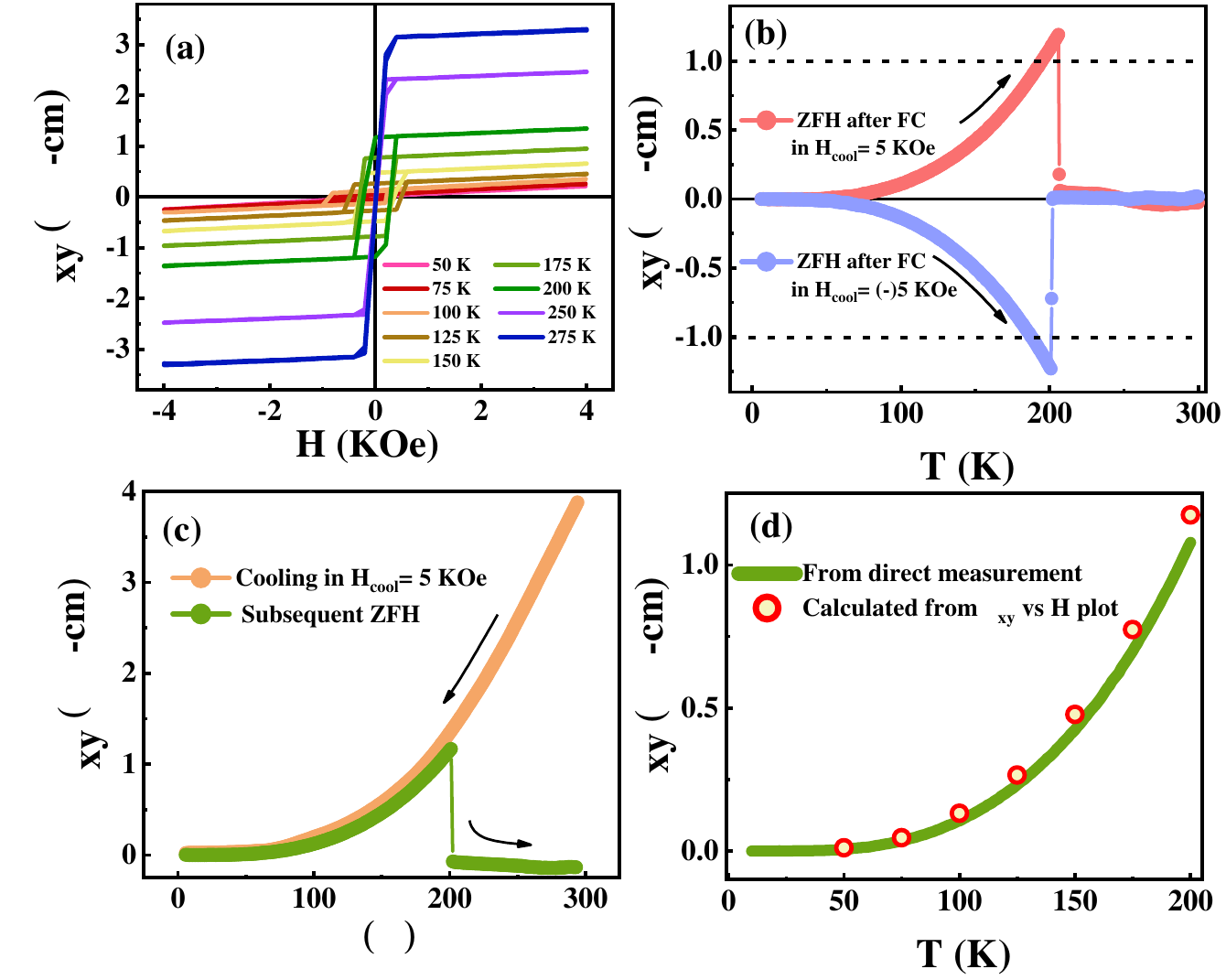}
	\caption{ (a) Field dependence of  Hall resistivity ($\rho_{xy}$) of TbMn$_6$Sn$_6$ at different temperatures. (b)$\rho_{xy}$ vs $T$ while zero field heating  after cooling in 5 kOe (red) and -5 kOe (blue) respectively. (c) Temperature dependence of Hall resistivity ($\rho_{xy}$ vs $T$ ) while heating in zero field after cooling in 5 KOe (symmetrized). (d) Comparison between $\rho_{xy}$ vs $T$ data obtained from direct measurement and $\rho_{xy}$ vs $H$ loops. The red circles represent the Hall voltage obtained from the $\rho_{xy}$ vs $H$ plot at different temperatures [Fig.~\ref{fig:TRM_Hall} (a)], while the green line represent the Hall voltage measured in zero field while heating after being field cooled in 5 kOe.}
	\label{fig:TRM_Hall}
\end{figure}

\label{sec:mag}
To characterize the magnetic nature of the crystal, we measured $M$ as a function of temperature ($T$) under $H$ = 1 kOe for both $H \parallel c$ and $H\perp c$ [Fig.~\ref{fig:MAG}(a)]. Clear signature of SR transition is seen around $T_{sr} \approx$ 310 K.  While  for $T > T_{sr} $, the easy axis lies perpendicular to the $c$ axis, it turns parallel to the $c$-axis below $T_{sr}$.
\par
We recorded isothermal $M$ versus $H$ curves with $H \parallel c$ at different temperatures [Fig.~\ref{fig:MAG}(b)]. At 50 K, the sample shows a rectangular hysteresis loop with a coercivity of 4.3 kOe. At higher temperature, the loop shrinks and almost vanishes above about 200 K. For $T < T_{sr}$, all the curves are characterized by a sharp rise of $M$ at low $H$ followed by saturation at higher fields. However, at  330 K ($> T_{sr}$), the sharp low field rise is absent, and $M$ increases sluggishly with $H$ and eventually tends to saturate above 30 kOe. Our observed magnetic data are at par with the previous results of a FIM ground state of the sample~\cite{RMS_singlecrystal_Jmmm,RMS_mag}.

\par
The most fascinating observation of our magnetization data is the  ultrasharp jump in the thermo-remanent magnetization (TRM). For TRM measurement, we first cooled the sample from 350 K in a field of $H_{cool}$ = 5 kOe down to 5 K. At 5 K, $H_{cool}$ was removed and a small field of 100 Oe was applied. The sample was subsequently heated in 100 Oe, and the thermo-remanent magnetization $M^{tr}$ was measured [Fig.~\ref{fig:TRM_mag} (a)]. The 100 Oe heating curve (red) follows the 5 kOe cooling curve (blue) till about 200 K, and beyond  200 K, $M^{tr}$ sharply falls to a value which is more than one order of magnitude less. On heating the sample beyond 200 K, it follows the regular $M$ vs $T$ plot recorded in 100 Oe of field, and it even shows the anomaly at $T_{sr}$. The jump in $M^{tr}$ is found to be rather sharp, and it occurs within 0.5 K of temperature. Notably, the signature of TRM and the jump in $M$ is absent when measured perpendicular to the $c$ axis (see Fig. 6 of  Appendix).
\par
We repeated the TRM measurements at different values of $H_{cool}$, and the measurement field during heating was also varied. Fig.~\ref{fig:TRM_mag} (b) shows the TRM data with $H_{cool}$ = 5 kOe and the subsequent heating in 20 Oe. Here a similar jump is observed close to 202 K. We have plotted the saturation magnetization, $M_{sat}$ (red triangles) obtained from $M$-$H$ loops alongside $M^{tr}$, and they match quite well. This indicates that there is a complete arrest of the spins due to field cooling, and $M^{tr}$ is as high as $M_{sat}$.
\par
To further confirm this unusual jump in the remanent magnetization, we cooled the sample in a negative field, $H_{cool}= -$ 5 kOe. While heating from 5 K in 100 Oe, $M$ remains negative up to about 200 K due to the obvious negative remanent $M$ due to cooling in a negative field. On reaching about 200 K, $M$ sharply jumps to a positive value. It appears that cooling in positive and negative fields produces TRM curves [Figs.~\ref{fig:TRM_mag} (a) and (c) respectively] that are mirror images of one another. We also measured TRM from two different batches (B1 and B2) of crystals, and the jump is present in both the batches (see Fig. 6 of Appendix).
\par
Fig.~\ref{fig:TRM_mag} (d) shows the TRM data measured for different $H_{cool}$, and in every value of the cooling field, the jump is present. The temperature value of the observed jump varies slightly around 200 K (within $\pm$ 2 K) for different sets of measurements, and it is not systematic. It is  to be noted that the values of $M$ in the heating path after being cooled in different $H_{cool}$ are almost independent of the cooling field. 

\par
TbMn$_6$Sn$_6$ is reported to show a large anomalous Hall effect~\cite{TMS_AHE1,TMS_AHE2}, which depends upon the magnetization of the sample. Since our magnetization measurements indicate the retention of $M$ up to about 200 K and a rapid fall beyond this temperature, it is essential to investigate the effect of TRM on the Hall voltage. First, we measured the standard field dependence of Hall resistivity $\rho_{xy}$ at various $T$ ranging from 50 K to 275 K as shown in Fig.~\ref{fig:TRM_Hall} (a). TbMn$_6$Sn$_6$ shows distinct anomalous Hall behavior above 50 K where $\rho_{xy}$ saturates after a certain value of applied field $H$ and the saturated value increases with increasing $T$. $\rho_{xy}$ shows hysteresis loops above 50 K, and the coercivity decreases to almost zero above 200 K. 
\par 
It is known that $\rho_{xy}$ as a whole is composed of the ordinary part ($\rho_{xy}^O$)and the anomalous part ($\rho_{xy}^A$) and can be written as~\cite{intrinsic_PRL}:
\begin{equation}
\centering
   \rho_{xy}=\rho_{xy}^O+\rho_{xy}^A = R_0H + 4\pi R_s M 
   \label{eqn:Hall}
\end{equation}
           
Here $R_0$ and $R_s$ are ordinary and anomalous Hall coefficients respectively. We calculated the value of $\rho_{xy}^A$  at different temperatures by finding the value of $\rho_{xy}$ at $H$ = 0 after linearly extrapolating the high-field region of the $\rho_{xy}$ vs $H$ curves~\cite{Hall_symm2}. At 275 K, $\rho_{xy}^A$ has a value of 3.23 $\mu\Omega$-cm, which is very close to the value reported in the literature~\cite{TMS_AHE3}.
\par
To check whether the exotic feature observed in our TRM data has any effect on Hall resistivity or not, we measured $\rho_{xy}$ with  similar cooling and heating protocols. We first cooled the sample in a field $H_{cool}$ down to 6 K, and subsequently heated it in zero field up to 300 K. During these heating and cooling, $\rho_{xy}$ was measured as a function of $T$. In Fig.~\ref{fig:TRM_Hall} (b), we have plotted the zero field heating curve of  $\rho_{xy}$ after being field cooled in $H_{cool}$ = 5 kOe and $-$5 kOe (red and blue curves respectively). Although the heating measurement is performed in zero field, the curves show substantial Hall resistivity with its value being  positive or negative depending upon the sign of $H_{cool}$. The Hall voltage thus observed is spontaneous in the absence of $H$, and it is associated with the remanent $M$ of the sample. We, therefore, call the observed $\rho_{xy}$ during zero-field-heating (ZFH) as thermo-remanent Hall (TRH), $\rho_{xy}^{tr}$. The sharp jump in $\rho_{xy}^{tr}$ around 200 K, concomitant with the jump observed in the TRM data, is present in both $H_{cool}$ = 5 kOe and $-$5 kOe curves, where the measured $\rho_{xy}^{tr}$ tends to attain zero value from positive and negative values respectively. Beyond the jump, $\rho_{xy}^{tr}$ continues to maintain a low value with increasing $T$.   
\par
In Fig.~\ref{fig:TRM_Hall} (c), we have plotted an antisymmetrized $\rho_{xy}$ versus $T$ plot. Temperature variation of the as-recorded $\rho_{xy}$ for $H_{cool} =\pm$ 5 kOe have been used to antisymmetrize the data. The green curve was recorded during cooling in 5 kOe field ($\rho_{xy}^{FC}$) and the orange curve ($\rho_{xy}^{tr}$) represents the TRH data.  Both $\rho_{xy}^{tr}$ and $\rho_{xy}^{FC}$ indicate the increase in $\rho_{xy}$ with increasing $T$, and they match quite well. However, they deviate from each other when the ZFH curve shows the sharp jump at around 200 K. In Fig.~\ref{fig:TRM_Hall} (d), we have plotted the anomalous Hall resistivity as a function of $T$ (red scattered points, $\rho_{xy}^A$) obtained from $\rho_{xy}$ vs $H$ loops [Fig.~\ref{fig:TRM_Hall} (a)] along with the ZFH curve, $\rho_{xy}^{tr}$. We find that both the data match surprisingly well, and it indicates that  $\rho_{xy}^{tr}$ primarily represents the AHE of the sample.
\par
Anomalous Hall resistivity, $\rho_{xy}^{A}$ is connected to longitudinal resistivity, $\rho_{xx}$ as:
\begin{equation}
  \rho_{xy}^{A} = \alpha\rho_{xx} + \beta \rho_{xx}^2 
  \label{eqn:scaling}
\end{equation}
We have plotted $\rho_{xy}^{tr}$ as a function of $\rho_{xx}$ in the inset of Fig.~\ref{fig:rho_M} ($\rho_{xx}$ vs. $T$ data is presented in Fig. 8 (a) of Appendix). The solid line is a fit to the data using eqn.~\ref{eqn:scaling}. Our fitting generates $\alpha$ = 8.6 $\times$ 10$^{-4}$ and $\beta$ = 1.37 $\times$ 10$^{-4}$ $(\mu\Omega-cm)^{-1}$. They are very close to the 
reported values of $\alpha$ and $\beta$ in the literature, and it indicates that the intrinsic contribution also dominates in $\rho_{xy}^{tr}$.
\par
From the present study, it is evident that TbMn$_6$Sn$_6$ retains its magnetization when the field-cooled sample is heated in a very low field up to a temperature as high as 200 K. The value of the TRM at a particular $T$ is the same as that of the saturation magnetization $M_s$. This large value of the retained TRM is responsible for the observed spontaneous Hall resistivity while heating in zero field. Further, we measured $\rho_{xx}$ vs $T$ in the field cooled state, where almost no effect of thermo-remanence is seen [see Fig.8(b) of Appendix], possibly because $M$ affects $\rho_{xy}$ directly as compared to its effect on $\rho_{xx}$.

\begin{figure}[hbt]
	\centering
	\includegraphics[width = 8 cm]{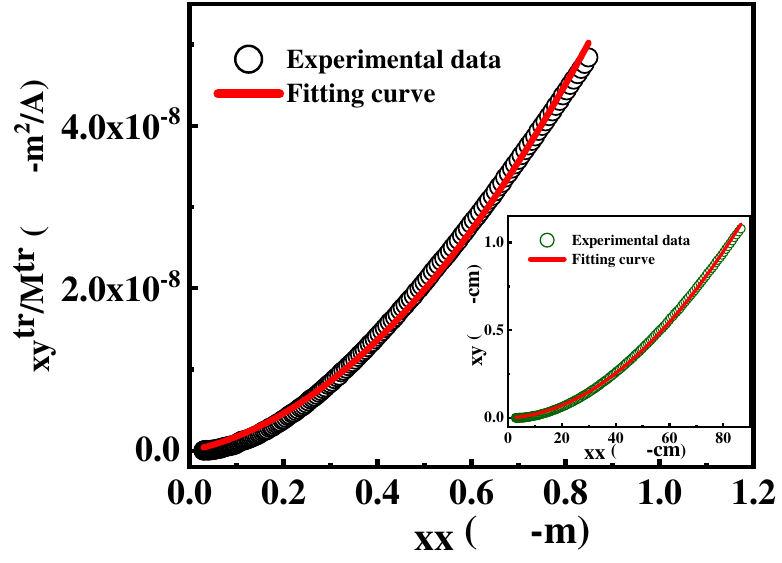}
	\caption{ $\rho_{xy}^{tr}/M^{tr}$ as a function of zero-field longitudinal resistivity $\rho_{xx}$. The solid line is a fit to the data using eqn.~\ref{eqn:rho_M}.}
	\label{fig:rho_M}
\end{figure}
\par
To check the role of TRM on $\rho_{xy}^{tr}$, we used the following relation connecting two quantities in the absence of applied $H$ (see eqn. (3) of Ref.~\cite{scale}). 
\begin{equation}
  \rho_{xy} = S_HM\rho_{xx}^2\left(1+\frac{\alpha^{\prime}}{\rho_{xx}}\right)
  \label{eqn:rho_M}
  \end{equation}
Here $S_H$ and $\alpha^{\prime}$ are parameters.
We have plotted $\rho_{xy}^{tr}/M^{tr}$ as a function of $\rho_{xx}$ in Fig.~\ref{fig:rho_M} for $T <$ 200 K, the curve is fitted using eqn.~\ref{eqn:rho_M}. The fitted curve traces the experimental data reasonably well, and it indicates that TRH and TRM vary conjointly. The values of parameters found from the fitting are, $S_H$= 0.056 V$^{-1}$, and $\alpha^{\prime}$ = 2.13 $\times$ 10$^{-7} \Omega$-m. 

\par
TRM, or development of remanent $M$ on field cooling is not uncommon, and it is observed in spin glass~\cite{nordblad}, magnetic glass showing first-order phase transition~\cite{chaddah,sbroy}, or in superparamagnet~\cite{superpara}. Recent magnetic study on TbMn$_6$Sn$_6$ has predicted cluster-glass-like feature due to the presence of competing exchange interactions~\cite{ANE_ATHE_AHE_nat}, and the observed TRM can be assigned to the glassy magnetic state and the presence of anisotropy.  The underlying mechanism for TRM in a spin glass or cluster glass is the existence of energy minimum in the free energy landscape, where the system gets arrested on field cooling. Below room temperature, TbMn$_6$Sn$_6$ shows large uniaxial anisotropy($ K_u \sim$ MJm$^{-3}$)~\cite{anisotrpy_2} with the $c$ axis being the easy axis. Cooling in a positive  $H_{cool} \parallel c$ produces favorably oriented spin clusters along the direction of $H$, and they remain frozen even when the field is removed and the sample is heated back. The spins only get de-arrested at higher $T$ ($\sim$ 200 K) when $K_u$ turns weak and/or the thermal fluctuation of the spins becomes strong. It is interesting to note that the coercivity of the magnetic hysteresis loop of TbMn$_6$Sn$_6$ vanishes above about 200 K~\cite{ANE_ATHE_AHE_nat,TMS_chern_Topology}, and a recent muon-spin rotation study indicates the development of magnetic fluctuation in the similar $T$-range~\cite{muSR}. The complete saturation of $M$ and $\rho_{xy}$, and a sharp jump of $M$ indicate that the spins behave cooperatively. This cooperative nature of the spin dynamics can be traced back to the strong intersite exchange interaction, $J\sim -$ 29 meV~\cite{exchange}.

\par
In conclusion, we observe a large TRM on field cooling, which the sample retains even when heated in zero or very low field and it eventually collapses above about 200 K. The same behavior is seen in our Hall measurement data, where we observe a zero-field spontaneous Hall effect on the field-cooled sample. Such thermo-remanence and the sharp jump observed in $M^{tr}$ and $\rho_{xy}^{tr}$ is rather exotic, and can be exploited for future spintronics and magnetic switching devices.

T.R. thanks the UGC, India  for the research assistance.  The UGC DAE-CSR Kolkata center, where the transport measurements were performed, is duly acknowledged.
\newpage
\appendix

\section{Sample preparation and characterization}
\begin{figure}[b]
	\centering
	\includegraphics[width = 8 cm]{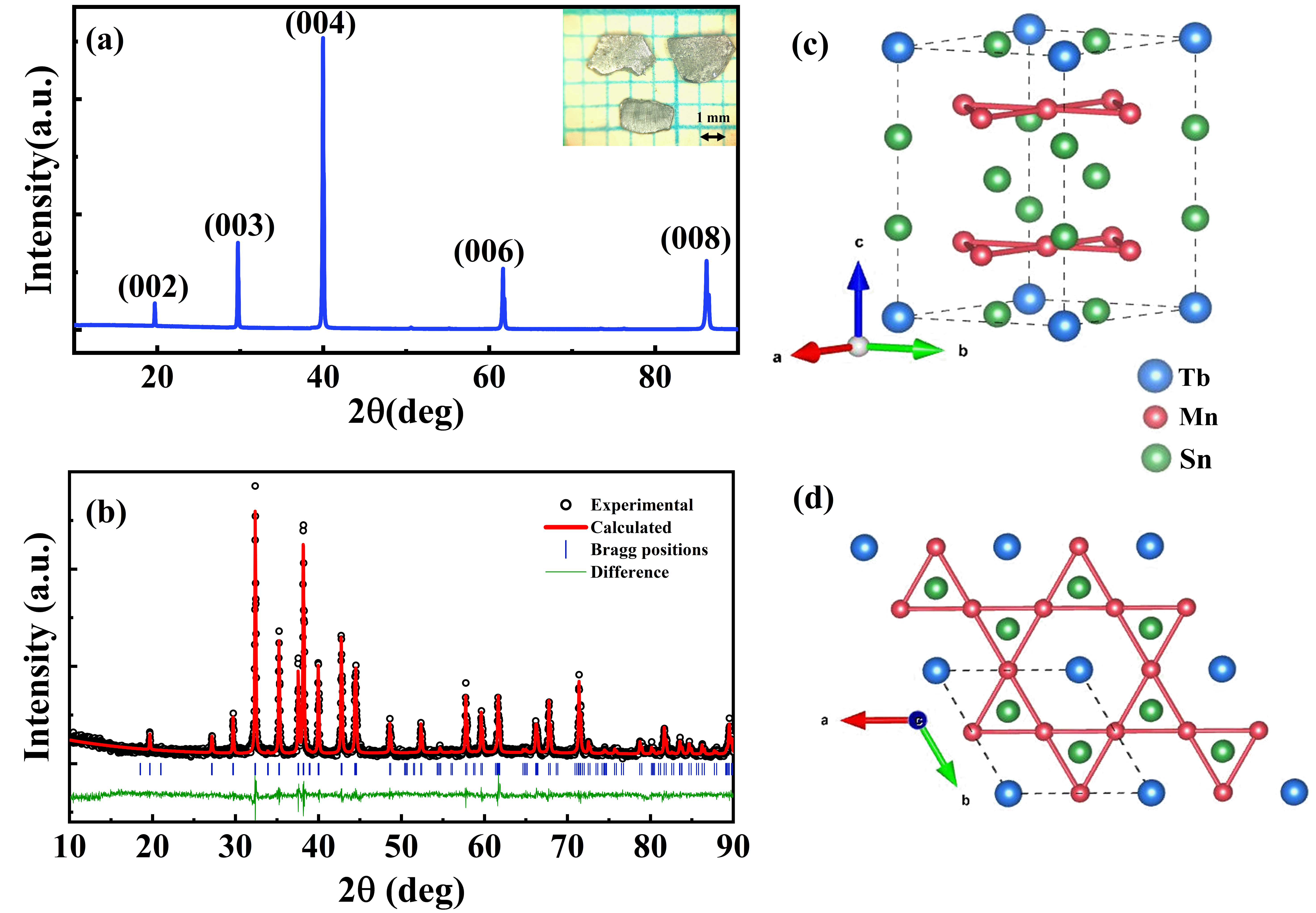} 
	\caption{(a) XRD pattern of a single crystal piece (from batch B1) at room temperature (b) Powder XRD pattern (symbols) with Rietveld refinement (solid line) at  room temperature for B1 crystals. Perspective views of the crystal structure of TbMn$_6$Sn$_6$ (c) unit cell along $c$ axis (d) ab-plane}
	\label{fig:xrd}
\end{figure}

High-quality single crystals of TbMn$_6$Sn$_6$ were synthesized by a Sn-flux method as described elsewhere~\cite{TMS_chern_Topology}. We prepared two different batches of samples, namely, B1 and B2 with the almost same heat treatments. The crystals obtained from both the batches were plate-like of size around a few mm with the $c$-axis perpendicular to the plane of the plate [see the inset of Fig.~\ref{fig:xrd} (a)]. This is evident from our $\theta$-2$\theta$ x-ray diffraction scan taken on the flat surface of the crystal, where we can see only the (00$l$) reflections  [Fig.~\ref{fig:xrd} (a)]. We also ground few crystals into powder and performed a powder x-ray diffraction (PXRD) as shown in Fig.~\ref{fig:xrd} (b) (for the B2 samples) along with Rietveld refinements  using MAUD software package~\cite{maud}. The sample crystallizes  in the hexagonal $P6/mmm$ structure with lattice constants, $a$ = 5.53~\AA, and $c$ = 9.02~\AA, which match well with the previous reports~\cite{Riberolles2022,Yin2020}. The perspective views of the crystal structure for two different orientations are shown in Figs.~\ref{fig:xrd} (c) and (d).

\begin{figure}[t]
	\centering
	\includegraphics[width = 8 cm]{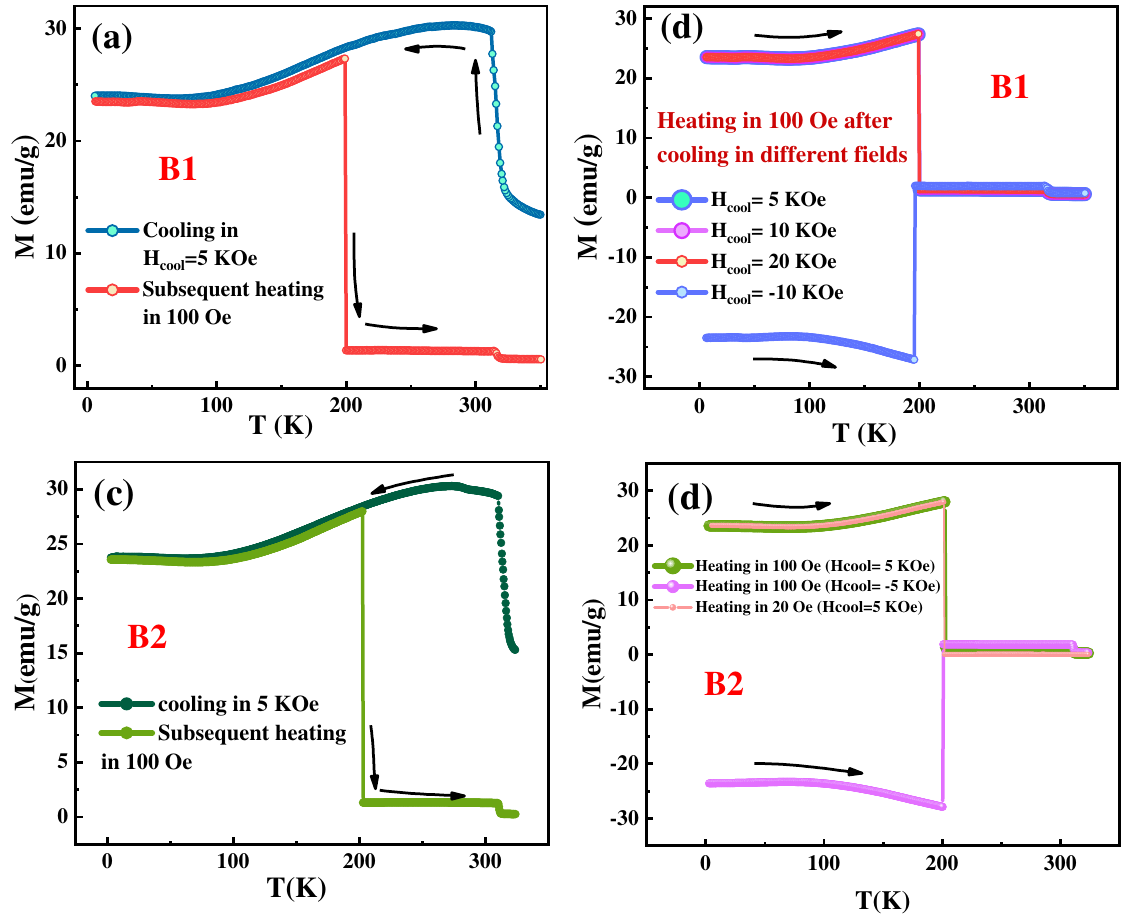}
	\caption{Temperature dependence of magnetization is shown for two batches of sample B1 and B2 (a) Cooling the sample (B1) in 5 kOe field and then heating in 100 Oe (b) heating curve of the sample (B1) in 100 Oe field after being field-cooled in different $H_{cool}$ (c) cooling the sample (B2) in 5 kOe and heating in 100 Oe and (d) heating curves of the sample (B2) after being field-cooled in different $H_{cool}$ .}
	\label{fig:mag_jump}
\end{figure}

\section{Measurement techniques}
Magnetic measurements were carried out using the vibrating sample magnetometer module of a commercial physical properties measurement system (PPMS, Quantum Design) as well as on a SQUID-VSM (MPM3) of Quantum Design. Hall and resistivity measurements were performed on a cryogen-free high magnetic field system (Cryogenic Ltd. UK). The magnetic and the Hall measurements were repeated for different crystal pieces obtained from B1 and B2 batches. We measured the Hall resistivity of the sample using the standard four-probe method applying the external field $H$ along the $c$-axis. To remove the longitudinal resistivity contribution coming from voltage probe misalignment, we antisymmetrize the data using the Hall resistivity measured for both positive and negative values of $H$, namely, $\rho_{xy} = \frac{1}{2}[\rho_{xy}(+H) - \rho_{xy}(-H)]$~\cite{Hall_symm1,Hall_symm2}.

\section{Magnetic data}
To check whether the jump in magnetization data was authentic, we prepared a second batch of the same sample following same synthesis protocol and repeated few measurements. We label the 1st batch of sample as B1 and second batch of sample as B2. In Fig.~\ref{fig:mag_jump}, we have plotted the field-cooled $M$ versus $T$ data for crystals from B1 and B2. One can see that the characteristic jump and thermo-remanent magnetization are identical in two different batches.

\par
\begin{figure}[t]
	\centering
	\includegraphics[width = 8 cm]{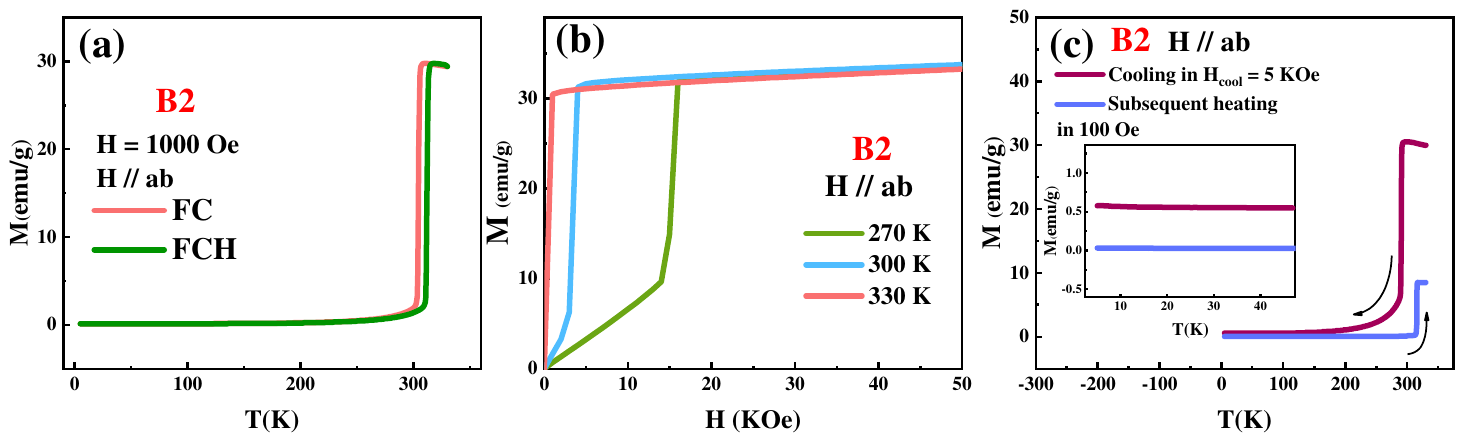}
	\caption{(a) Temperature dependence of magnetization with $H \perp c$. There is a small thermal hysteresis near the spin reorientation transition temperature probably due to the transition is of 1st order. (b) Field dependence of magnetization at three different temperatures  $T$= 270 K, 300 K, 330 K. (c) we tried to check whether there is some thermo-remanence in magnetization along ab plane also so the sample was cooled in $5$ kOe field and then heated in $100$ Oe field. No signature of jump due to thermo-remanence was found.}
	\label{fig:mag_ab}
\end{figure}

We also measured magnetization in the same field-cooled heating protocol with $H \perp c$ as shown in Fig.~\ref{fig:mag_ab}. The sample shows metamagnetic transition in this orientation below the spin reorientation temperature. We also studied $M$ vs $T$ by cooling in $H_{cool}$ = 5 kOe and heating in 100 Oe [Fig.~\ref{fig:mag_ab} (c)]. We see taht the field cooled-heating data do not follow the field-cooled heating curve indicating the absence of thermo-remanet magnetization in the $ab$ plane. Also, no jump in the $T$ variation of the remanent $M$ is observed in this direction. It indicates that TRM and the sharp jump is only present when $M$ is measured along the $c$ direction.

\begin{figure}[t]
	\centering
	\includegraphics[width = 8 cm]{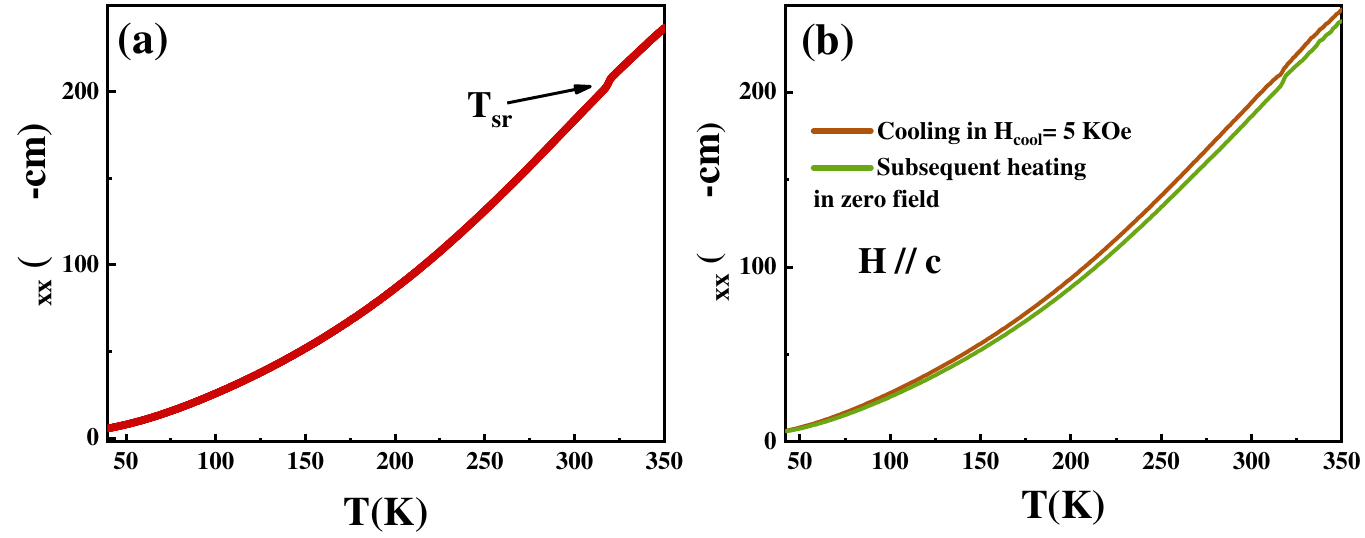}
	\caption{ (a) The longitudinal resistivity ($\rho_{xx}$) as a function of temperature up to $ 350 K$ on crystal from batch B1. (b) Temperature dependence of $\rho_{xx}$ while cooling in 5 kOe and then heating the sample in zero field in a similar manner we measured $\rho_{xy}$ vs $T$. We applied both positive and negative 5 kOe fields as the cooling field and obtained the final curve after symmetrization using the formula $\rho_{xx} = \frac{1}{2}[\rho_{xx}(+H) + \rho_{xx}(-H)]$.}
	\label{fig:res}
\end{figure}

\section{Transport data}
Longitudinal resistivity, $\rho_{xx}$ is plotted as a function of temperature in Fig.~\ref{fig:res} (a). The sample shows a metallic behavior. At around 315 K, a little hump is seen in the $\rho_{xx}$ vs $T$ curve representing the spin-reorientation transition. Field-cooling and subsequent zero-field heating do not show much difference indicating that that there is no signature of thermo-remanence in the magneto-resistance data.  
\newpage
\bibliographystyle{apsrev4-2} 
\bibliography{mycite}

\end{document}